\documentclass[twocolumn,showpacs,preprintnumbers,amsmath,amssymb]{revtex4}

\usepackage{graphicx}
\usepackage{dcolumn}
\usepackage{bm}

\begin{document}

\title{Spin coherence of holes in GaAs/AlGaAs quantum wells}

\author{M. Syperek$^{1,2}$, D. R. Yakovlev$^{1,\dag}$, A. Greilich$^{1}$, J. Misiewicz$^{2}$, M. Bayer$^{1}$,
D. Reuter$^{3}$, and A.~D. Wieck$^{3}$}

\affiliation{$^1$Experimentelle Physik II, Universit\"at Dortmund,
D-44221 Dortmund, Germany} \affiliation{$^2$ Institute of Physics,
Wroc{\l}aw University of Technology, 50-370 Wroc{\l}aw, Poland}
\affiliation{$^3$Angewandte Festk\"orperphysik, Ruhr-Universit\"at
Bochum, D-44780 Bochum, Germany}

\date{\today}

\begin{abstract}
The carrier spin coherence in a p-doped GaAs/(Al,Ga)As quantum
well with a diluted hole gas has been studied by picosecond
pump-probe Kerr rotation with an in-plane magnetic field. For
resonant optical excitation of the positively charged exciton the
spin precession shows two types of oscillations. Fast oscillating
electron spin beats decay with the radiative lifetime of the
charged exciton of 50 ps. Long lived spin coherence of the holes
with dephasing times up to 650 ps. The spin dephasing time as well
as the in-plane hole g factor show strong temperature dependence,
underlining the importance of hole localization at cryogenic
temperatures.
\end{abstract}

\pacs{42.25.Kb, 78.55.Cr, 78.67.De}
\maketitle

Recently the investigation of the coherent spin dynamics in
semiconductor quantum wells (QW) and quantum dots has attracted
much attention, due to the possible use of the spin degree of
freedom in novel fields of solid state research such as spin-based
electronics or quantum information processing ~\cite{Awschalom,
Zutic,Vincenzo}. Until now the interest has been mostly focused on
the spin coherence of electrons, while experimental information
about the spin coherence of holes is limited \cite{Marie}. The
hole as a Luttinger spinor has properties strongly different from
the electron spin, such as a strong spin-orbit coupling, a strong
directional anisotropy, etc. It plays an important role also in
coherent control of electron spins, since in many optical schemes
charged electron-hole complexes are proposed as intermediate
manipulation states \cite{Imamoglu}.

Earlier, the hole spin dynamics in GaAs-based QWs has been
measured by optical orientation detecting photoluminescence (PL)
either time-integrated or time-resolved
\cite{Damen,Roussignol,Baylac,Baylac2,Marie}. Experimental studies
addressed the longitudinal spin relaxation time $T_{1}$
\cite{Damen,Roussignol,Baylac} and the dephasing time $T^*_{2}$,
exploiting the observation of hole spin quantum beats
\cite{Marie}. The reported relaxation times vary from 4 ps
\cite{Damen} up to $\sim$1 ns \cite{Baylac,Marie} demonstrating
strong dependence on doping level, doping density and excitation
energy. A major drawback of PL techniques is, however, that the
spin coherence can be traced only as long as both electrons and
holes are present and photoluminescence can occur. Further, they
work only for studying the spin dynamics of minority carriers in a
sea of majority carriers and are therefore restricted to undoped
or n-type doped QWs. However, then the holes can interact with
electrons, providing additional relaxation channels via exchange
or shake-up processes \cite{Baylac,Uenoyama}. These mechanisms can
be excluded for p-doped structures if the hole spin relaxation
occurs on time scales longer than the radiative annihilation of
electrons. A pump-probe Kerr rotation (KR) technique using
resonant excitation allows to realize such measurements, which up
to now have been reported only for bulk p-type GaN \cite{Hu} and
not yet for low-dimensional systems.

The theoretical analysis of the hole spin dynamics in QWs has been
focused on free holes \cite{Meier,Uenoyama,Ferreira,Lu} by
considering different relaxation mechanisms: (i) a Dyakonov-Perel
like mechanism, (ii) an acoustic phonon assisted spin-flip due to
spin mixing of valence band states, (iii) an exchange induced
spin-flip due to scattering on electrons, which resembles the
Bir-Aronov-Pikus mechanism, but for holes. Recently the attention
has been drawn on the role of hole localization and the dephasing
caused by fluctuations of the in-plane g factor has been
calculated \cite{Semenov}.

In this paper we use time-resolved pump-probe Kerr rotation
\cite{Baumberg1994} to investigate the spin coherence of holes in
a p-doped GaAs/Al$_{0.34}$Ga$_{0.66}$As single QW with a low hole
density. We find spin dephasing times reaching almost the ns-range
at a temperature $T=1.6$~K with a hole in-plane g factor of about
0.01. Both quantities decrease strongly with increasing
temperature, suggesting the important influence of hole
localization. We discuss also a mechanism that provides generation
of spin coherence for the hole gas under resonant excitation of
the positively charged exciton.

\begin{figure}
  \centering
\includegraphics[scale=1.5]{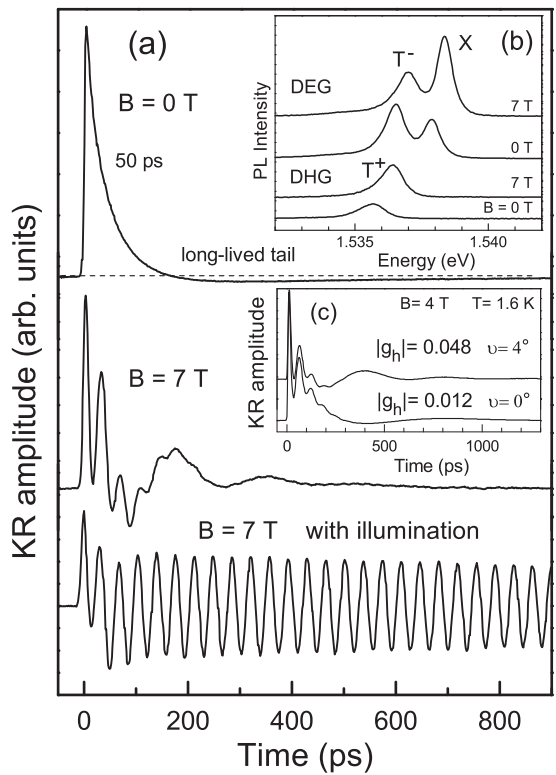}
\caption{ (a) KR traces for a p-type 15 nm-wide
GaAs/Al$_{0.34}$Ga$_{0.66}$As QW vs time delay between pump and
probe pulses at $B=0$ and 7~T with field tilted by
$\vartheta=4^{\circ}$ out of QW plane. Laser at 1.5365 eV is
resonant with T$^{+}$ line. Power was set to  5 and 1 W/cm$^{2}$
for pump and probe, respectively. Bottom trace was recorded with
additional laser illumination at 2.33~eV. $T=1.6$~K. (b) PL
spectra for DHG (excitation at 1.579~eV) and DEG regime (above
barrier excitation at 2.33~eV). (c) Comparison of KR traces for
$\vartheta=0$ and 4$^{\circ}$.} \label{fig1}
\end{figure}

The structure was fabricated by molecular-beam epitaxy on a (100)
oriented GaAs substrate. A 15 nm-wide GaAs QW was grown on top of
a 380 nm-thick Al$_{0.34}$Ga$_{0.66}$As barrier and overgrown by a
190 nm-thick Al$_{0.34}$Ga$_{0.66}$As layer. 21 nm-thick layers
with Al$_{0.34}$Ga$_{0.66}$As effective composition realized by
GaAs/AlAs short-period superlattices were deposited on both sides
of the QW in order to improve interface planarity. Two
$\delta$-doped layers with Carbon acceptors were positioned
symmetrically at 110 nm distance from the QW. The hole gas
concentration and mobility in the QW are $1.51 \times 10^{11}$
cm$^{-2}$ and $1.2 \times 10^{5}$ cm$^{2}$/Vs, respectively, as
determined by Hall measurements at $T=4.2$~K. It was possible to
deplete the hole density in the QW by above barrier illumination
and even invert the majority carrier type, resulting in a diluted
electron gas. The sample temperature was varied from 1.6 to 6~K.

A mode-locked Ti:Sapphire laser with a repetition rate of 75.6 MHz
and a pulse duration of $\sim$1.5 ps ($\sim$1 meV full width at
half maximum) was used for optical excitation. The laser beam was
split into a circularly polarized pump and a linearly polarized
probe beam. Both beams where focused on the sample surface to a
spot diameter of $\sim$100 $\mu$m. Magnetic fields $B \leq$ 10 T
were applied about perpendicular to the structure growth $z$-axis
(Voigt geometry). In a pump-probe KR experiment the pump pulse
coherently excites carriers with spins polarized along the $z$
axis. The subsequent coherent evolution of the spins in form of a
precession about the magnetic field is tested by the probe pulse
polarization. To detect the change of the linear probe
polarization plane (the KR angle), a homodyne technique based on
phase-sensitive balanced detection was used.

Photoluminescence spectra excited above and below the band gap of
the Al$_{0.34}$Ga$_{0.66}$As barriers are shown in Fig.~1(b) at
$B=0$ and 7~T. A single PL line corresponding to the positively
charged trion T$^+$ (consisting of two holes and one electron) is
seen for the regime of diluted hole gas (DHG) established for
below-barrier excitation. After inverting the type of majority
carriers to the DEG regime by above barrier illumination the PL
spectra consist of the exciton (X) and negatively charged trion
(T$^-$) lines.

The type of majority carriers in the QW can be identified by the
KR signals measured at $B=7$~T, with the laser energy tuned to the
trion resonance. The bottom trace in Fig.~1(a) was measured with
additional above-barrier illumination (DEG regime) and shows
long-lived electron spin beats with a dephasing time of 2.5~ns
which is considerably longer than the radiative decay time of
resonantly excited trions of about 50~ps. The precession frequency
corresponds to a g factor $\mid g_{e} \mid = 0.285\pm0.005$, which
is typical for electrons in GaAs-based QWs.

Without above-barrier illumination in the DHG regime, fast
electron precession is observed only during $\sim$ 200 ps after
pump pulse arrival [see middle trace in Fig.~1(a)]. This signal is
caused by the coherent precession of the electron in T$^+$ and
disappears with the trion recombination. The electron beats are
superimposed on the hole beats with a much longer precession
period. The hole beats decay with a time constant of about 100~ps
and can be followed up to 500~ps delay. At these long times the KR
signal is solely given by coherent hole precession.

Experimentally, it is difficult to observe the hole spin quantum
beats due to the very small in-plane hole g factor. To enhance the
visibility we tilted the magnetic field slightly out of the plane
by an angle $\vartheta=4^{\circ}$ to increase the hole g factor by
mixing the in-plane component ($g_{h,\perp}$) with the one
parallel to the QW growth axis ($g_{h,\parallel}$), which
typically is much larger: $g_{h}(\vartheta) =
\sqrt{g^2_{h,\parallel} \sin^2 \vartheta + g^2_{h,\perp} \cos^2
\vartheta}$ \cite{gfactor}. The strong change of the hole beat
frequency with the tilt angle is seen in Fig.~1(c). The precession
frequency is analyzed by $\omega_h = \mu_B \mid g_{h} \mid B /
\hbar$, where $\mu_B$ is the Bohr magneton, and gives $\mid
g_{h,\perp} \mid = 0.012\pm0.005$ for $\vartheta=0^{\circ}$ and
$\mid g_{h} \mid = 0.048\pm0.005$ for $\vartheta=4^{\circ}$.

\begin{figure}
  \centering
\includegraphics[scale=1]{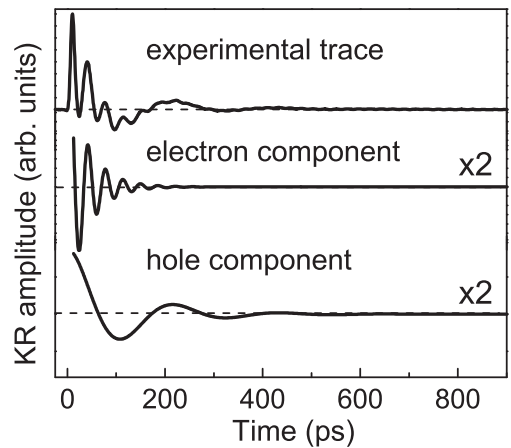}
\caption{Top trace: KR signal measured at $B= 7$~T for
$\vartheta=4^{\circ}$. Bottom traces are obtained by separating
electron and hole contributions (see text). Excitation conditions
as in Fig.~1.} \label{fig2}
\end{figure}

The electron and hole contributions to the KR amplitude,
$\Theta_{\textrm{K}}$, can be separated by fitting the
experimental data with a superposition form of exponentially
damped harmonic functions for electrons and holes:
\begin{equation}\label{formula1}
    \Theta_{\textrm{K}}=\sum_{i=e, h} A_i \exp(-\frac{\Delta t}{T_{2,i}^*})\cos(\omega_i \Delta
    t).
\end{equation}
$A_i$ are the corresponding signal amplitudes at pump-to-probe
delay $\Delta t=0$, and $T_{2,i}^*$ are the spin dephasing times.
An example for a decomposition of the KR signal in the DHG regime
is shown in Fig.~2.

\begin{figure}
  \centering
\includegraphics[scale=1]{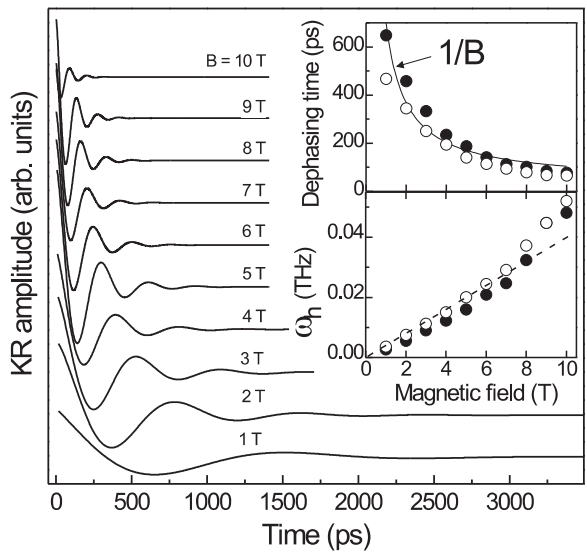}
\caption{Hole component of KR signal at different magnetic fields
and $\vartheta=4^{\circ}$. Top inset: Magnetic field dependence of
the hole dephasing time $T_2^*$. Solid line is $1/B$ fit to data.
Bottom inset: Hole spin precession frequency vs magnetic fields.
Line is guide to the eye. In inserts closed and open circles show
the data measured for pump to probe powers of 1 to 5~W/cm$^{2}$
and 5 to 1~W/cm$^{2}$, respectively. $T=1.6$~K.} \label{fig3}
\end{figure}

Let us turn now to the hole coherence. Figure 3 shows the hole
contribution to the KR signal for different $B$ at $T=1.6$~K. From
the fit by Eq.(1) we have obtained the dephasing time $T_2^*$,
which is plotted versus $B$ in the inset. A very long lived hole
spin coherence with $T_2^*=650$~ps is found at $B=1$~T. With
increasing $B$ up to 10~T it shortens to 70 ps. The field
dependence can be well described by a $1/B$-form (see the line in
the inset), from which we conclude that the dephasing shortening
arises from the inhomogeneity of the hole g factor $\Delta
g_h=0.0007$ in the QW, which is translated into a spread of the
precession frequency: $\Delta \omega_h = \Delta g_h \mu_B B /
\hbar$.  Since $T_2^{\star} \propto 1 / \Delta \omega_h$, this
explains the $1/B$-dependence of the dephasing time. The magnetic
field dependence of the hole precession frequency in the lower
inset of Fig.~3 shows an approximate $B$-linear dependence up to 7
T. For higher fields a super linear increase is seen which
indicates a change of the hole g factor due to mixing between
heavy and light hole states, induced by the field.

Two sets of experimental data measured for pump to probe powers 1
to 5~W/cm$^{2}$ and 5 to 1~W/cm$^{2}$ are compared in the insets
of Fig.~3. The very similar results demonstrate performance of the
experiment in the linear regime for both pump and probe beams with
power not exceeding 5~W/cm$^{2}$.

\begin{figure}
  \centering
\includegraphics[scale=1]{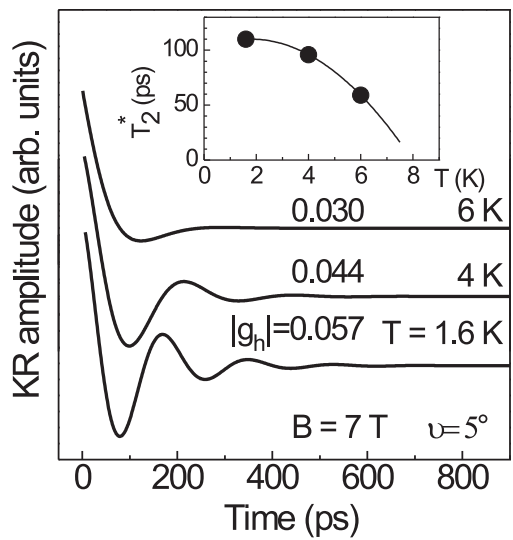}
\caption{Temperature dependence of the hole KR signal at $B=7$~T
and $\vartheta=5^{\circ}$.  Pump and probe powers are set to 1 and
5~W/cm$^{2}$, respectively. Inset: Hole spin dephasing time
$T_2^*$ vs temperature.} \label{fig4}
\end{figure}

Insight into the origin of the long hole spin coherence can be
taken from KR at varying temperatures. The data in Fig.~4 measured
at $\vartheta = 5^{\circ}$ show that (i) the dephasing time
$T_2^{\star}$ decreases from 110 down to 60~ps when increasing the
temperature from 1.6 to 6~K (see also inset), and (ii)
simultaneously the precession frequency decreases notably
corresponding to a g factor decrease from 0.057 to 0.030.

These results can be naturally explained by hole localization in
the QW potential relief due to monolayer well width fluctuations.
The localization energy does not exceed 0.5~meV, which is
comparable with the thermal energy at $T=6$~K. Free holes are
expected to have a short spin coherence time $T_2$ limited by the
efficient spin relaxation mechanisms due to the spin-orbit
interaction \cite{Uenoyama,Ferreira,Lu}. For localized holes these
mechanisms are mostly switched off and one can expect long $T_2$
times. However, in a KR experiment we do not measure the $T_2$
time, but rather the ensemble dephasing time $T_2^*$
(Ref.~\onlinecite{Semenov}), as confirmed by the $1/B$ dependence
in the inset of Fig.~3. The $T_2^*$ time gives a lower boundary
for the spin coherence time $T_2$. Therefore we can conclude, that
the $T_2$ for localized holes is at least 650~ps. Thermal
delocalization of holes on the one hand decreases the role of
inhomogeneities and reduces $\Delta g_h$, which should lead to
longer dephasing times. On the other hand, then the fast
decoherence of free holes becomes the limiting factor for the spin
beats dynamics.

Mixing of heavy and light hole states in a QW is enhanced by
localization effects. This should be detectable by an increase of
the in-plane hole g factor, which is close to zero for free holes
\cite{Marie,Semenov,Winkler}. The decrease of the hole g factor
with increasing temperature shown in Fig.~4 is consistent with a
hole delocalization scenario.

We turn now to discussing the mechanism for optical generation of
hole spin coherence in a QW with a DHG. The generation mechanism
is similar to the one suggested for singly charged quantum dots
\cite{Shabaev,Gre_06} and QWs with a DEG \cite{Kennedy}. In our
experiment pump and probe are resonant in energy with the
positively charged trion T$^{+}$. Due to the considerable
heavy-light hole splitting, the circularly polarized pump creates
holes and electrons with well-defined spin projections, $J_{h,z} =
\pm 3/2$ and $S_{e,z} = \pm 1/2$, respectively, according to the
optical selection rules \cite{Meier}. Therefore, $\mid \Uparrow
\Downarrow \downarrow \rangle$ ($\mid \Uparrow \Downarrow \uparrow
\rangle$) trions can be generated by a $\sigma^+$ ($\sigma^-$)
polarized pump. Here the thick and thin arrows give the spin
states of holes and electrons, respectively.

The pump pulse duration is much shorter than the spin coherence
and the electron-hole recombination times. If in addition the pump
duration is shorter than the charge coherence time of the trion
state the pulse creates a coherent superposition of a resident
hole from the DHG and a hole singlet trion T$^{+}$. The spin state
of the resident hole with arbitrary spin orientation before
excitation can be described by $\alpha \mid \Uparrow \rangle +
\beta \mid \Downarrow \rangle$, where $|\alpha|^2+|\beta|^2=1$.
Without magnetic field and for fields oriented normal to the
$z$-axis, the net spin polarization of the hole ensemble is zero,
so that the ensemble averaged coefficients are equal:
$\overline{\alpha} = \overline{\beta}$.

For $\sigma^+$ polarized excitation, for which injection of an
$\mid \Uparrow \downarrow \rangle$ electron-hole pair is possible,
the excited superposition is given by $\alpha \mid \Uparrow
\rangle + \beta \cos (\Theta/2) \mid \Downarrow \rangle + i \beta
\sin (\Theta/2) \mid \Uparrow \Downarrow \downarrow \rangle$. Here
$\Theta = \int {\bf d} \cdot {\bf E} (t) dt / \hbar$ is the
dimensionless pulse area with the pump laser electric field ${\bf
E} (t)$ and the dipole transition matrix element ${\bf d}$. In
general, the hole-trion superposition state may be driven
coherently by varying the pulse area, giving rise to
Rabi-oscillations as reported recently for (In,Ga)As quantum dots
\cite{Gre_06}. Such oscillations have not been found yet in QWs,
most probably due to the fast carrier dephasing, in particular for
strong excitation. Dephasing of the superposition occurs shortly
after the pulse on a time scale of a few ps, converting the
coherent polarization into a population consisting of holes with
original spins $\Uparrow$ and $\Downarrow$ and trions with
$\Uparrow \Downarrow \downarrow$.

In a simplified picture, the spin coherence generation can be
described as follows:  The $\sigma^+$ polarized pump creates with
certain efficiency trions T$^{+}$ of  spin configuration $\mid
\Uparrow \Downarrow \downarrow \rangle$. By this process $\mid
\Downarrow \rangle$ holes are pumped out of the DHG, leaving
behind holes with opposite spin $\mid \Uparrow \rangle$. Right
after the pump pulse the KR signal is contributed by the $\mid
\Uparrow \rangle$ hole from the DHG and $\mid \downarrow \rangle$
electron of the T$^{+}$. The further evolution of the coherent
signal depends on the strength of external magnetic field applied
perpendicular to the $z$-axis.

At $B$=0, the carrier spins experience no Larmor precession. The
electron spin relaxation time usually exceeds the lifetime of
trions, which is limited by radiative decay, by one-two orders of
magnitude. Trion recombination returns the hole to the DHG with
the same spin orientation as it was pumped out, if no electron
spin scattering occurred in the meantime. This compensates the
induced spin polarization and nullifies the KR signal at delays
exceeding the trion lifetime. Indeed, the KR signal in the top
trace in Fig.~1(a) shows a fast decay with a time constant of
$\sim$50 ps, which is characteristic for radiative trion
recombination in GaAs/(Al,Ga)As QWs \cite{Finkelstein}. The
long-lived tail of the signal has a very small amplitude and is
due to hole coherence provided by weak spin relaxation of
electrons in T$^{+}$ and/or hole relaxation in the DHG during the
trion lifetime.

In finite magnetic fields, the carrier spins start to precess
about $B$. Due to the electron spin precession in T$^{+}$, the
hole spin returned to the DHG after trion recombination will not
compensate the spin polarization of the resident holes. Therefore,
a long-lived hole coherence with considerable amplitude will be
induced. This coherence is observed in the KR signal as  spin
beats with low frequency (see Figs. 1 and 3). Note that the Larmor
precession of the resident holes may also contribute to generation
of hole spin coherence, but the effect is proportional to the
ratio of the hole and electron Larmor frequencies and therefore
will be rather small.

Let us compare the spin coherence generation for QWs with DHG and
DEG resonantly excited in the T$^{+}$ and T$^{-}$ states,
respectively. We are interested in a long-lived spin coherence
which goes beyond the trion lifetime, i.e. in spin coherence
induced for the resident carriers. In both cases the amplitude of
the KR signal is controlled by the ratio of the electron spin beat
period to the trion lifetime. Nevertheless, the two cases are
quite different as for DHG the precessing electron is bound in the
T$^{+}$ trion, while for DEG the background electron precesses. In
the latter case the electron precession in T$^{-}$ is blocked due
to the singlet spin character of the trion ground state.

To conclude, a long-lived spin coherence has been found for
localized holes in a GaAs/(Al,Ga)As QW with a diluted hole gas.
The spin coherence time exceeds 650 ps and is still masked by the
spin dephasing due to g factor inhomogeneities. Localization of
holes suppresses most spin relaxation mechanisms inherent for free
carriers. It is also worth to note, that due to the p-type Bloch
wave functions the holes do not interact with the nuclear spins,
which provides the most efficient spin relaxation mechanism for
localized electrons \cite{Merk02}.

{\bf Acknowledgements.} This work was supported by the BMBF
program 'nanoquit'.



\begin{references}
\bibitem{gfactor} The value $\mid g_{h,\parallel} \mid = 0.60\pm0.01$ was determined from the Zeeman splitting
of PL lines at $B=7$~T applied along the QW growth axis.
\end{references}
\end{document}